\documentclass[aps,prl,amsmath,amssymb,secnumarabic,floatfix,twocolumn,superscriptaddress]{revtex4-2}
\usepackage{graphicx}
\usepackage{amsmath}
\usepackage{amssymb}
\usepackage{color}
\usepackage{soul}
\usepackage{float}
\usepackage{longtable}
\begin{document}

\title{Lattice dynamics and elastic properties of black phosphorus}%

\author{Eva A. A. Pogna$^{1,2,*}$, Alexe{\"i} Bosak$^{3*}$, Alexandra Chumakova$^{3}$, Victor Milman$^{4}$, Bj{\"o}rn Winkler$^{5}$, Leonardo Viti$^{1}$, Miriam S. Vitiello$^{1}$}

\affiliation{NEST, CNR-Istituto Nanoscienze and Scuola Normale Superiore, P.zza S. Silvestro 12, 56127 Pisa, IT\\
$^{2}$Istituto di Fotonica e Nanotecnologie, Consiglio Nazionale delle Ricerche, P.zza L. da Vinci 32, 20133 Milano, Italy\\
$^{3}$European Synchrotron Radiation Facility, BP 220, F-38043 Grenoble Cedex, France\\
$^{4}$Dassault Syst\'emes BIOVIA, Cambridge, CB4 0WN United Kingdom\\
$^{5}$Institute of Geosciences, Goethe University Frankfurt, Altenh{\"o}ferallee 1, 60438 Frankfurt am Main, Germany\\
* equally contributing authors\\
evaariannaaurelia.pogna@cnr.it
}
\begin{abstract}
We experimentally determine the lattice dynamics of black phosphorus
layered crystals through a combination of x-ray diffuse scattering and
inelastic x-ray scattering, and we rationalize our experimental
findings using \textit{ab initio} calculations. From the phonon
dispersions, at terahertz frequencies, we derive the full
single-crystal elastic tensor and relate it to the macroscopic elastic response of black phosphorus, described by the elastic moduli. The elastic stiffness coefficients obtained here, provide an important
benchmark for models of black phosphorus and related materials, such
as phosphorene and black phosphorus nanotubes, recently emerged quantum materials, disclosing a huge potential for nanophotonics,
optoelectronics and quantum sensing.
	 
\end{abstract}

\maketitle
\section{Introduction}
Black phosphorus (bP) is the most thermodynamically stable allotrope
of phosphorus under ambient
conditions\cite{Akahama1987,Clark2010}. First synthesized in 1914 by
Bridgman \cite{Bridgman1914}, it has recently gained renewed
interest\cite{Xia2014,Ling2015} as a layered two dimensional quantum
material appealing for nano-optoelectronic\cite{Li2014black,Zhu2015}
and nano-photonic\cite{Deng2018,Zhou2017} applications, and as a core
material for the synthesis of single-layer phosphorene\cite{Xia2014}.
bP has promising semiconducting properties, including a near-infrared
direct band gap (E$_g$= 0.3 eV in the bulk\cite{Warschauer1963}) and
high room-temperature hole mobility ($>$650 cm$^2$V$^{-1}$s$^{-1}$ in
the bulk\cite{Morita1986} and $>$1,100 cm$^2$V$^{-1}$s$^{-1}$ in 15 nm
thick films\cite{Xia2014}), which have allowed the demonstration of
high-performance devices for electronics (\textit{e.g.} $\sim$10$^5$ on/off
ratio field-effect transistors\cite{Ma2017}), optoelectronics
(\textit{e.g.} photodetectors at telecom\cite{Yuan2015},
mid-infrared\cite{Guo2016,Chen2017} and terahertz (THz)\cite{Viti2015,Viti2016,Viti2017} frequencies) and photonics
(\textit{e.g.} saturable absorbers\cite{Sotor2015,Luo2015}).  The peculiar
layered structure of bP not only allows for easy exfoliation to the
few-layers limit\cite{Liu2014,Brent2014,Yasaei2015,Kang2015}, but is also characterized by strong in-plane anisotropy, which is reflected in the
electronic band structure of bP\cite{Asahina1984}, in its optical properties
(\textit{e.g.} thickness and polarization dependent
absorption\cite{Xia2014,Low2014,Li2015} and linear
dichroism\cite{Xia2014,Qiao2014}), in its electrical properties
(\textit{e.g.} mobility and effective mass\cite{Xia2014}) and thermal
properties\cite{Jang2015}.  The inherent anisotropy of bP results in
orthogonal prominent directions for electron and heat transport, maximizing charge mobility or thermal conductivity,
respectively\cite{Zhang2020}.  Its optical anisotropy has been
exploited to design room-temperature THz frequency
photodetectors with controlled detection dynamics \cite{Viti2016},
anisotropic perfect absorbers\cite{Xiong2017}, tunable
polarizers\cite{Qing2018,Shen2018}, synaptic devices\cite{Tian2016}
and even to explore novel device functionality via strain engineering
of bP phonon modes\cite{Wang2015Remarkable}.

Recently, the growth of ultrathin bP on the centimeter scale\cite{Wu2021} has opened a route for the development of
bP-based wafer-scale architectures and improved the technological
relevance of bP. However, despite the continuously growing interest in
this material, the experimental characterization of its elastic
properties is still rather incomplete. The off-diagonal elements of bP elastic tensor have not been experimentally determined yet, and the available measurements of the diagonal elements, by ultrasound (US)\cite{Yoshizawa1986,Kozuki1991} and inelastic neutron scattering (INS)\cite{Fujii1982}, are inconsistent with atomistic model calculations\cite{Appalakondaiah2012,Wang2015}
(see Table I, with determinations of the coefficient $C_{11}$ differing by up to 100 GPa).
This severely limits the quantum engineering of bP-based devices,
especially in the THz-regime, the frontier region of the
electromagnetic spectrum.
Moreover, a detailed study of the lattice dynamics of bP is still missing, with measurements limited to zone-center optical phonons by Raman
spectroscopy\cite{Ribeiro2018,Kim2015,Wu2015identifying}, which proved
to be a powerful tool for identifying crystal orientation exploiting
the dependence of the Raman fingerprints on the relative
orientation between the excitation polarization and bP
crystallographic axes.

Here, we apply x-ray diffuse scattering (DS)\cite{Bosak2015} and inelastic x-ray scattering (IXS)\cite{Baron2016} to determine the energy dispersion of the THz phonons of orthorhombic bP throughout
its Brillouin zone. DS measurements allow a fast mapping of
reciprocal space, and are very efficient for identifying the regions of
interest for IXS measurements, which probe the collective motions of atoms with high energy and momentum
resolution. From the
retrieved phonon dispersions, we derive all the elastic stiffness coefficients, providing an experimental benchmark
for parameter-free model calculations of the key electronic parameters
bP and bP-based low dimensional materials.
\onecolumngrid
\begin{center}
	\begin{figure}[h]
		\includegraphics[]{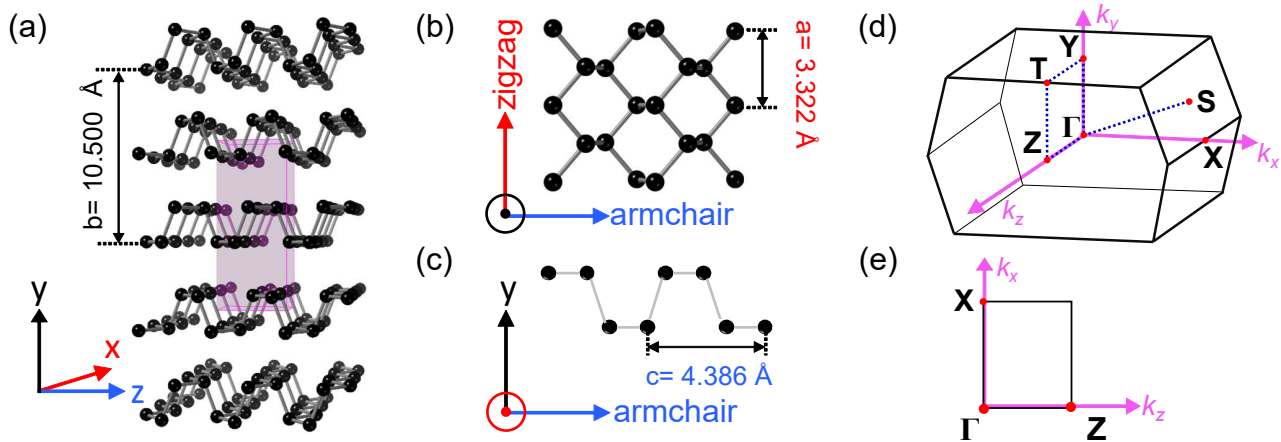}
		\caption{(a), (b), and (c) Atomic structure of A17 face-centered
			orthorhombic bP, showing the primitive unit cell
			(pink shadowed cell) described by optimized lattice
			parameters $a$ = 3.322 \AA, $b$ = 10.500
			\AA, $c$ = 4.386 \AA, along the x,y and z axes
			respectively. In this paper, x corresponds to the zigzag direction
			and z corresponds to armchair direction; (d) and (e) Brillouin zone
			and high-symmetry points (red circles) of bulk bP
			(d) and corresponding two-dimensional (2D) surface (e),
			\textit{i.e., phosphorene}. The blue dotted lines in (d)
			indicate the directions explored by IXS
			measurements.}
		\label{f:structure}
	\end{figure}
\end{center}
\twocolumngrid
\section{RESULTS AND DISCUSSION}
Black phosphorus has an A17 face-centered orthorhombic structure
(see Fig.\ref{f:structure}) belonging to the Cmce space group and
D$^{18}_{2h}$ point group, which includes a twofold in-plane rotational
symmetry. In contrast, other group V elements (\textit{e.g.} Sb,
As and Bi) crystallize in A7 rhombohedral structure, which P adopts only
in high-pressure conditions, above 5 GPa\cite{Scelta2017}.
Orthorhombic bP has a reduced symmetry compared with its elemental layered
material counterparts from group IV, such as graphene, silicene and
germanene\cite{balendhran2015}, which have a hexagonal lattice with
sixfold in-plane rotational symmetry (D$^{4}_{6h}$ point group).  In
each layer of bP, the $sp^3$ hybridization of P atoms leads to the
formation of a puckered honeycomb structure featuring strong covalent
bonds, while the layers are stacked together by weak van der Waals
forces.

The C-centered unit cell of bP, in Fig.\ref{f:structure} a-c, is described by the lattice parameters $a$ = 3.322 \AA, $b$ = 10.500 \AA, and $c$ = 4.386 \AA, consistently
with previous measurements $a\sim3.3$ \AA,
b$\sim10.5$ \AA{} and c$\sim{4.4}$ \AA{} by x-ray powder
diffraction\cite{Brown1965,Henry2020} and scanning transmission
electron microscopy\cite{Wu2015}.  In the physical coordinate system, the axis
\textbf{x} is along the zigzag direction [100], \textbf{y} along the
stacking direction [010] and \textbf{z} is along the armchair direction
[001].  
The Brillouin zone and high-symmetry points in the reciprocal space of
face-centered orthorhombic bP are displayed in
Fig.\ref{f:structure}d-e. The primitive cell contains four atoms, such
that bP has 12 phonon branches: 3 acoustic and 9 optical.

High quality, single crystals of bP are obtained using a chemical
vapor transport method\cite{Khurram2020}, placing the evacuated quartz
tube containing red phosphorus into a double-zone tube furnace with
hot and cold ends at 600$^{\circ}$C and 500$^{\circ}$C, respectively.
Micro-Raman characterization is reported in the
Supplemental Material (SM)\cite{SM}.
The sample for DS and IXS measurements is laser cut to a cylinder
shape with 150 $\mu$m diameter and 50 $\mu$m height, cleaned by ethanol and
scotch tape cleavage. While the sample has a mosaic composition with
angular spread $\Delta\theta<1^{\circ}$, single grains with rocking
curve $\Delta\theta<0.1^{\circ}$ can be localized, providing the
required high crystalline quality for resolving the lattice dynamics
along the different crystallographic directions, even in the proximity
of the $\Gamma$ point.

DS maps are collected at the x-ray diffractometer\cite{Girard2019} of
the beamline ID28 of the European Synchrotron Radiation Facility
(ESRF), keeping the sample in nitrogen flux at room temperature. The
measurements are performed in shutterless mode using a monochromatic
beam with wavelength $\lambda$= 0.784 \AA{}, and energy $E$=15.814 keV,
focused on the sample over a 40$\mu$m-diameter spot. DS frames are
obtained by rotating the sample through 360$^{\circ}$ range,
orthogonally to the incoming beam, with an angular slicing of
0.25$^{\circ}$.
Figure 2 reports the experimental DS patterns as reconstructed high
symmetry reciprocal space planes, showing strong anisotropy of diffuse
features. All the DS features in the experimental maps are reproduced, in the shape and relative intensities, by ab initio calculations of the thermal diffuse scattering (TDS) by lattice vibrations (for details see SM and Refs. \cite{Clark2005,Perdew1996,Lejaeghere2016,Monkhorst1976,Grimme2010,Appalakondaiah2012}), indicating that the absence of other contributions. The TDS is highly dominated by soft (low-energy) phonons,
specifically by transverse acoustic modes propagating along the
stacking direction [010], particularly the one polarized along the
armchair direction.

To determine the energy dispersion associated with the energy-integrated
features in the DS maps, we perform an IXS experiment using the x-ray spectrometer\cite{Krisch2007} of
the beamline ID28 of ESRF. We operate with $\lambda$=0.6968 \AA{}
($E$= 17.794 keV) beam, providing an energy resolution of 3 meV, and focused on the sample over a 30 $\mu$m-diameter spot.
\onecolumngrid
\begin{center}
	\begin{figure}
		\includegraphics[width=\textwidth]{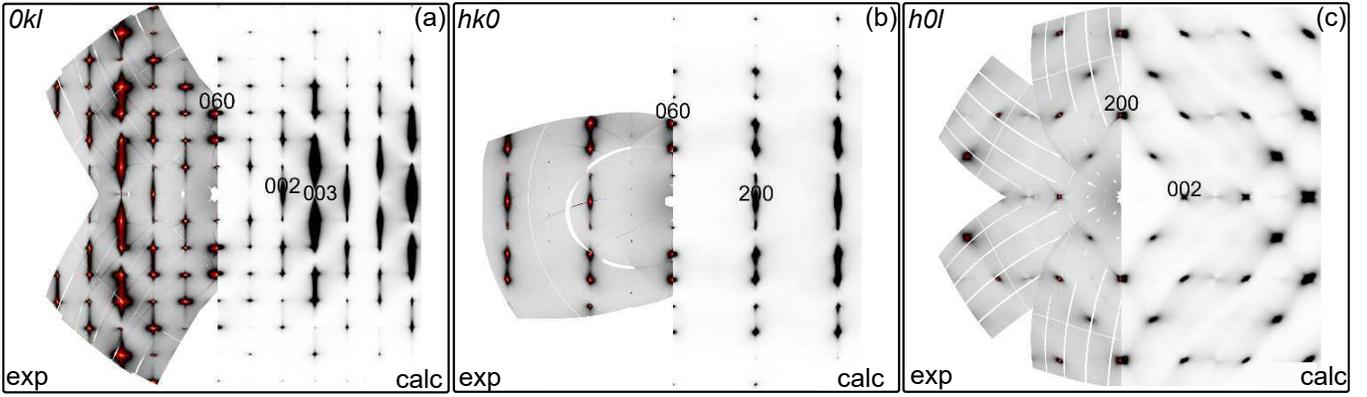}
		\caption{Reconstructed reciprocal space maps of bP compared with the \textit{ab initio} calculations of thermal diffuse scattering of (a) 0kl, (b) hk0 and (c) h0l planes. Laue symmetry elements are applied to remove the gaps due to the space between detector elements. Combined linear grayscale and logarithmic color scale visualization is applied to the experimental data as implemented in Albula software\cite{dectris}.}
		\label{f:diffuse}
	\end{figure}
\end{center}
\twocolumngrid
\newpage

\onecolumngrid\
\begin{center}
	\begin{figure}[h]
		\includegraphics[]{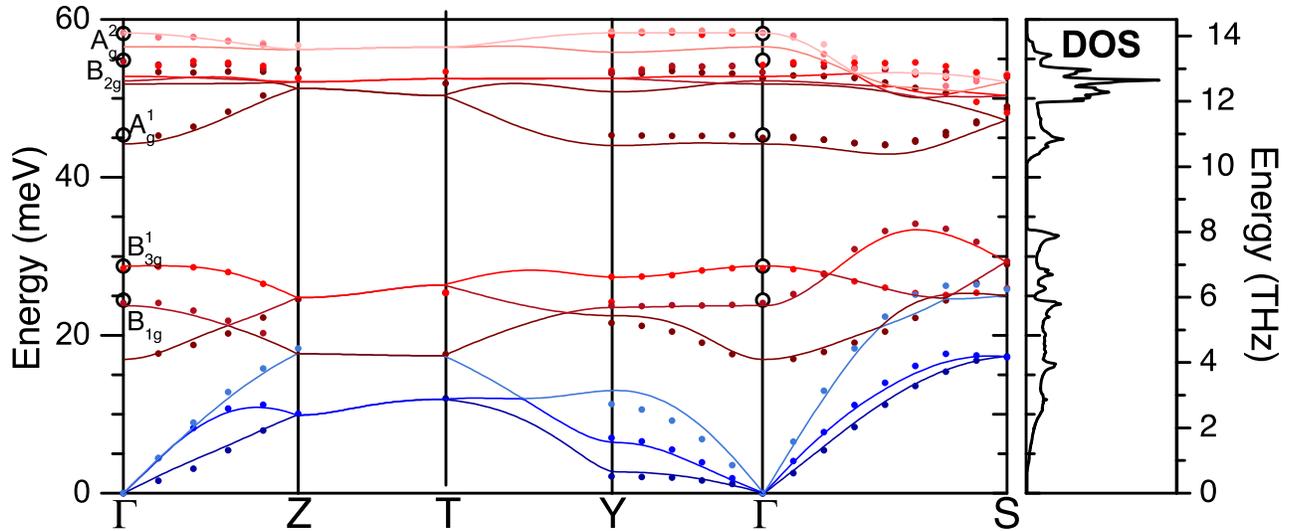}
		\caption{Experimental phonon dispersions of bP measured by IXS (symbols), compared with the predictions by \textit{ab initio} calculations (solid lines), together with the calculated phonon density of states (DOS, right panel). Black open circles correspond to Raman spectroscopy data measured on a twin sample reported in the SM, which are in good agreement with Ref.\cite{Ribeiro2018}.}
		\label{f:dispersion}
	\end{figure}
\end{center}
\twocolumngrid\
Given that bP phonons can approach 60 meV (\textit{i.e.} 15 THz), and, considering the small size of available high-quality crystals, IXS appears to be the ideal method to investigate the phonon
dispersion relations\cite{Bosak2006,Bosak2007}.
By analyzing the energy position of the inelastic peaks in the IXS spectra, we determine the energy dispersion of the active bP phonon modes along the $\Gamma$-Z, $\Gamma$-Y, and $\Gamma$-S directions, see
Fig.\ref{f:dispersion}. 
The analysis of the calculated eigenvectors indicates that in the $\Gamma$-Z direction, in the longitudinal geometry (00\textit{l} line), the scattered intensity is continuously transferred from the longitudinal phonon (\textit{i.e.} for 002+$\delta$) to the transverse phonon (\textit{i.e.} for 003+$\delta$). This exchange corresponds to the transition from the diffuse cloud around the strong 002 reflection to the bow-tie cloud centered on the weak 003 node in the TDS map of the 0kl plane in Fig.\ref{f:diffuse}a,c, and it has been previously observed in $\alpha$-quartz \cite{Bosak2018}.

The \textit{ab initio} calculations of phonon dispersions allow us
to optimize the intensity and contrast with the neighboring phonons
for a large selection of individual branches, making the experiment
time-efficient despite the large number of branches.  The \textit{ab
  initio} calculations are
indeed in good agreement with IXS data (see Fig.~\ref{f:dispersion}). Hence, they are used to derive the dispersions along the unexplored directions, and to evaluate the phonon density of states (DOS).
The DOS of bP, in Fig.~\ref{f:dispersion}, exhibits a
phonon gap between 8.7 and 10.8 THz, and the highest peaks close to the Raman mode B$_{2g}$, due to the flatness of the optical modes'
dispersion around the $\Gamma$ point. From the DOS, we extract the temperature dependence of a number of thermodynamic properties of bP, including entropy, enthalpy, free energy and heat capacity, see SM and Refs.\cite{Stephenson1969,Paukov1965,Schlesinger2002}.

In order to extract the elastic tensor $C_{\rm ij}$, we collect the
phonon dispersions with a fine step ($<$1 nm$^{-1}$) along:
\textit{(i)} high symmetry directions, for the diagonal
components $C_{\rm ii}$; \textit{(ii)} intermediate directions in high
symmetry planes, for the off-diagonal components ($C_{\rm ij}$
with i$\neq$j). Relevant sound velocities are obtained by fitting the dispersions and taking the slope at the zero momentum
limit. The energy correction for the analyzer opening (momentum
resolution) was introduced iteratively based on refined elastic
moduli.
\onecolumngrid\
\begingroup \squeezetable
\begin{table}[h]
	\caption{Elastic stiffness coefficients of bP in GPa as derived from the IXS data, compared to \textit{ab initio} calculations (Calc.) of bP\cite{Appalakondaiah2012} and phosphorene \cite{Wang2015}, and previously published data (Exp.) from INS\cite{Fujii1982} and US\cite{Yoshizawa1986,Kozuki1991}.}
	\centering
	\begin{ruledtabular}
		\begin{tabular}{c c c c c c c c c c}
			& C$_{11}$ & C$_{22}$ & C$_{33}$ & C$_{44}$ & C$_{55}$ & C$_{66}$ & C$_{12}$ & C$_{13}$ & C$_{23}$
			\\ [1ex]
			\hline
			Expt.(this paper). & 193 & 53 & 61 & 7 & 72 & 27 & 3 & 50 & -7\\
			
			INS Expt.\citenum{Fujii1982} & 250.1 & 70.6 & 57.4 & 7.8 & 57.4 & 21.3/26.1 & & &		 \\
			US Expt.\citenum{Yoshizawa1986}& 284 & 57 & 80 & 10.8 & 59.4 & 17.2 & & &
			\\
			US Expt.\citenum{Kozuki1991}& 178.6 & 53.6 & 55.1 & 5.5 & 14.5/15.6 & 21.3/11.1
			\\
			Calc.(this paper)& 186.0 & 70.4 & 54.6 & 6.3 & 63.3 & 22.7 & 10.3 & 41.8 & 1.2
			\\ 
			Calc.\citenum{Appalakondaiah2012}& 191.9 & 73.0 & 52.3 & 8.8 & 63.6 & 25.5 & 8.3 & 40.8 & 0.6
			\\
			Calc.\citenum{Wang2015}& 186.7 & & 46.5 && 39.8 && & 32.7 &\\
		\end{tabular}
	\end{ruledtabular}
	\label{table:Table1}
\end{table}
\endgroup
\twocolumngrid
In Table I, we compare the nine symmetrically independent non-zero components of the
elastic tensor of orthorhombic bP derived from the IXS, with \textit{ab initio} simulations for bulk
bP\cite{Appalakondaiah2012} and phosphorene\cite{Wang2015}), and with the available
values from INS\cite{Fujii1982} and US
measurements\cite{Yoshizawa1986,Kozuki1991}. 
The elastic stiffness coefficients determined here satisfy the stability conditions for orthorhombic crystals and are all positive except for $C_{23}$. The value retrieved for $C_{11}$ is significantly (by up to 90 GPa) lower than earlier measurements\cite{Yoshizawa1986,Fujii1982}, and 15 GPa larger than the US value from Ref.\cite {Kozuki1991}. The large discrepancy, with and amongst previous determinations, is due to the experimental limitations of US and INS, severely hindered by the availability of large high-quality samples, and highlights the unique opportunity offered by IXS to measure the phonon dispersions of bP.
US gives access to phonon modes of smaller wave vectors; however is more sensitive to defects than IXS, being a macroscopic method. Moreover, IXS overcomes the inherent limitations of INS related to sample size (a few millimeters), energy transfer and momentum resolution, enabling one to probe the entire Brillouin zone with a micrometer-size sample, for which high-crystal quality can be guaranteed.

Remarkably, the elastic stiffness coefficients from IXS, agree well with the density functional theory (DFT) calculations, which are evaluated from stress-strain relations, \textit{i.e.} not from the slope of the acoustic phonons.
An agreement within few GPa between experiment and DFT results, such as the one achieved with IXS data, is to be expected and testifies to the accuracy and robustness of the present determination.
The best estimate of the elastic tensor can be derived by averaging over selected values, \textit{i.e.} excluding values which are in clear disagreement with the others, see SM for a graphical comparison. This yields $C_{11}$ = 193 GPa, $C_{22}$ = 63 GPa, $C_{33}$ = 57 GPa, $C_{44}$ = 8 GPa, $C_{55}$ = 63 GPa, $C_{66}$ = 24 GPa, $C_{12}$ = 7 GPa, $C_{13}$ = 44 GPa, and $C_{23}$ = 1 GPa, for which we expect an error equal to 10\% on the diagonal terms and 5 GPa for the off-diagonal terms. The IXS values for $C_{12}$
and $C_{23}$ are the least reliable, since, being small by modulus, they
are evaluated using significantly larger input values of $C_{ii}$.

From the elastic stiffness coefficients, we can evaluate the
Voigt-Reuss-Hill average\cite{Voigt1928,Reuss1929,Hill1952} of bulk modulus $B\sim$34 GPa
and shear modulus $G\sim$26 GPa (details in the SM), providing an
estimation of the elastic response under hydrostatic pressure and upon
shape change, respectively.
The strong anisotropy of bP is testified by the distinct in-plane
Young's moduli along the armchair and zigzag directions equal to
$\sim$49 and $\sim$158 GPa, respectively, as derived from
the elastic tensor.  These values also provide approximate upper and lower
limits for phosphorene and phosphorene-based nanotubes, whose
controlled synthesis has been recently demonstrated\cite{Cai2017}.
Their mechanical properties are directly linked to the elasticity of
bP.
The value of $C_{11}$= 193 GPa provides the
upper estimate for the on-axis Young's modulus of homogeneous bP
nanotubes to be compared with the much larger value estimated for the
single-walled carbon nanotube (SWNT), which varies from 1.0 to
1.25-1.28 TPa\cite{Bosak2007}.  Very different mechanical and thermal
properties can be expected by changing the diameter
and bonding direction (armchair or zigzag)\cite{Wang2018}.

From the retrieved elastic stiffness coefficients, we can also extract
the Poisson's ratio $\nu$ (see SM and Ref.\cite{Boulanger1998}), which, for uniaxial strain along the \textbf{x} direction parallel to the pucker, takes a negative value $\nu_{yx}\sim$ -0.27 in the out-of-plane direction \textbf{y}, in
agreement with recent predictions for phosphorene\cite{Jiang2014}.
\begin{center}
	\begin{figure}[h]
		\includegraphics[width=0.5\textwidth]{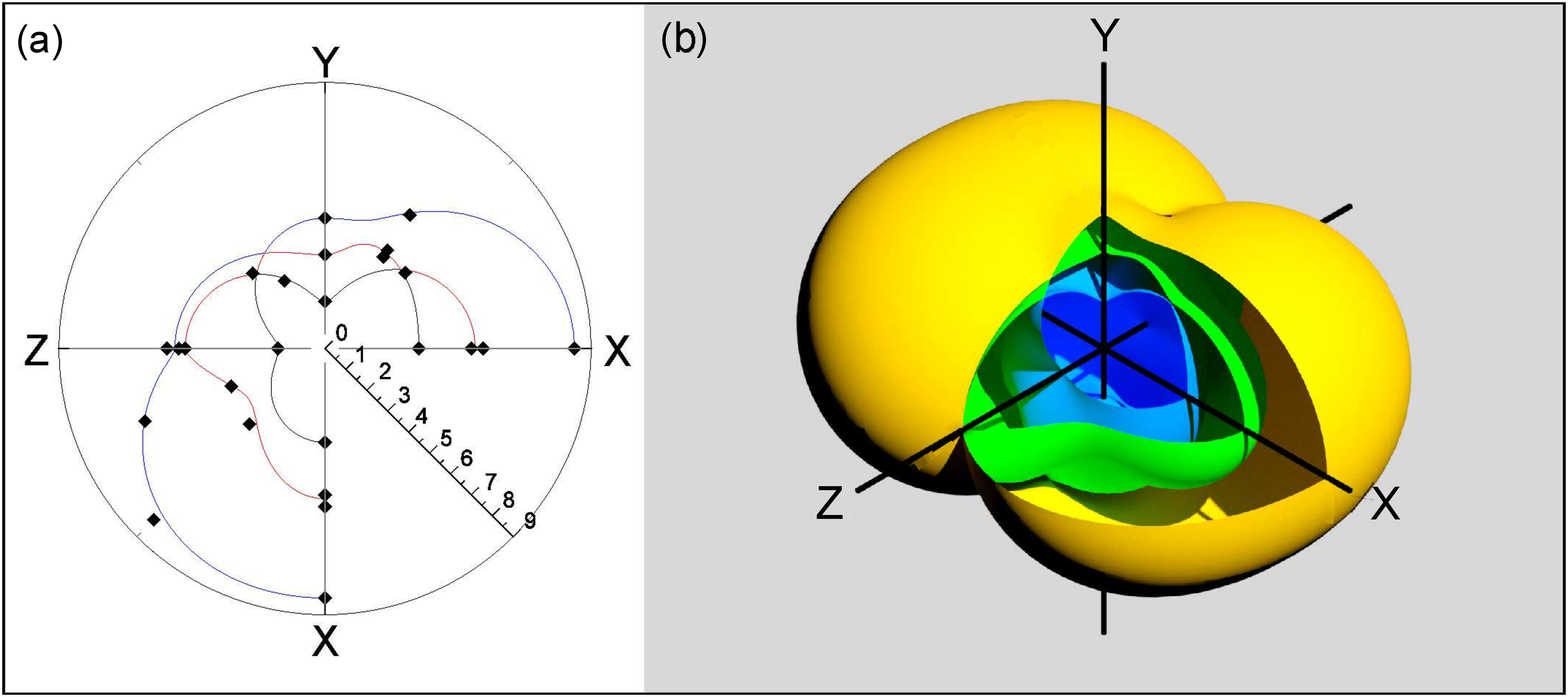}
		\caption{Directional distribution of sound velocities in bP in km/s: (a) Angular plot for the principal high symmetry planes. Symbols denote IXS-derived sound velocities; solid lines are calculated from Christoffel's equation\cite{Fedorov2013} with corresponding parameters. (b) 3D representation of sound velocities; colors correspond to the different eigenvalues of Christoffel's equation, sorted in increasing order.}
		\label{f:velocity}
	\end{figure}
\end{center}
The degree of anisotropy can be quantified by the shear anisotropic
factors for the $\{$100$\}$ plane $A_1$= 0.2, for $\{$010$\}$ plane $A_2$=
2.1, $\{$001$\}$ plane $A_3$= 0.4, these values differ from the unity value
expected for isotropic crystals (see SM and Ref.\cite{Liu2017}). The anisotropy can also be quantified by the universal anisotropy index \cite{Ranganathan2008} A$^u$= 3.75, which accounts for both shear and bulk contributions, and is zero for locally isotropic crystals.

The stiffness tensor allows generalizing the Hooke's law in three dimensions relating strains and stresses in the elastic regime, describing static deformations. Moreover, it encodes the propagation of sound waves within the material. Using the elastic stiffness coefficients in Table I, the sound waves velocities in bP are predicted by solving the Christoffel's equation \cite{Fedorov2013}, which allows one to determine both the dispersion relation of monochromatic density waves, \textit{i.e.} phase velocity, and the group velocity, using the derivatives and Hessian of the Christoffel matrix \cite{Jaeken2016}.
Angular plots of velocity, based on Christoffel's equation, in high
symmetry planes and in 3D, are given in
Fig.\ref{f:velocity}. Remarkably, in the $\Gamma$-Z direction, the
velocity of one of the two transverse waves is higher than that of the
longitudinal wave, meaning that, in the proximity of Z direction, the
outer surface of the sound velocity corresponds to the transverse wave.

This work provides the experimental determination of the full elastic tensor of orthorhombic
bP, achieved by measuring the phonon dispersions along the principal high-symmetry directions by IXS. The experimental elastic stiffness coefficients provide an important benchmark for atomistic model
calculations, which can then be used for the prediction of further properties of bP. The accurate (within a few GPa) and complete characterization of the
elasticity of bP allows us to correlate its microscopic properties (interatomic interactions)
with the macroscopic mechanical response and crystal stability, as described by elastic moduli and anisotropy indices,
providing a solid base for designing bP based devices with \textit{ad hoc} properties.
\newline

\begin{acknowledgments}
This work is supported by the European Research Council through the ERC Consolidator Grant (681379) SPRINT from the European Union through the EU Horizon 2020 research and innovation programme Graphene Flagship (GrapheneCore3), H2020-MSCA-ITN-2017 TeraApps (765426). BW is grateful for support from the BIOVIA Science Ambassador program.
\end{acknowledgments}

\end{document}